\def\aap{{\em Astronomy and Astrophysics}}

\def\araa{{\em Annual Review of Astronomy and Astrophysics}}
\def\apj{{\em The Astrophysical Journal}}
\def\apjs{{\em The Astrophysical Journal supplement}}
\def\apjl{{\em The Astrophysical Journal Letters}}

\def\grl{{\em Geophys. Res. Lett.}}
\def\jgr{{\em J. Geophys. Res.}}

\def\lrsp{{\em Living Reviews in Solar Physics}}

\def\nat{{\em Nature}}

\def\nu{{\em Nuclear Fusion}}

\def\prl{{\em Phys. Rev. Lett.}}

\def\pop{{\em Phys. Plasma}}
\def\pof{{\em Phys. Fluid}}

\def\physrep{{\em Physics Reports}}

\def\rspta{{Philosophical Transactions of the Royal Society of London Series A}}
\def\sci{{\em Science}}
\def\ssr{{\em Space Sci.~Rev.}}
\def\solphys{{\em Solar Physics}}
\def\sjetp{{Soviet Journal of Experimental and Theoretical Physics}}

\documentclass[a4paper]{jpconf}
\usepackage{graphicx}
\usepackage{caption} 
\begin{document}
\title{How Nanoflares Produce Kinetic Waves, Nano-Type III Radio Bursts, and Non-Thermal Electrons in the Solar Wind}

\author{H. Che}
\address{University of Maryland, College Park, MD, 20742, USA}
\address{NASA Goddard Space Flight Center, Greenbelt, MD, 20771, USA}

\begin{abstract}
Observations of the solar corona and the solar wind discover that the solar wind is unsteady and originates from the impulsive events near the surface of the Sun's atmosphere. How solar coronal activities affect the properties of the solar wind is a fundamental issue in heliophysics. We report a simulation and theoretical investigation of how nanoflare accelerated electron beams affect the kinetic-scale properties of the solar wind and generate coherent radio emission. We show that nanoflare-accelerated electron beams can trigger a nonlinear electron two stream instability, which generates kinetic Alfv\'en and whistler waves, as well as a non-Maxwellian electron velocity distribution function, consistent with observations of the solar wind. The plasma coherent emission produced in our model agrees well with the observations of Type III, J and V solar radio bursts. Open questions in the kinetic solar wind model are also discussed.
\end{abstract}
\section{Introduction}
\vspace*{0.5cm}
The origin of the solar wind is one of the most important unsolved problems in heliophysics. The concept of solar wind and the first steady hydrodynamic solar wind model were proposed by Parker in 1958 \cite{parker58apj}. The Parker model describes the solar wind as a continuous plasma outflow from the solar corona, maintained by the stationary expansion of the Sun's atmosphere. As pointed out by Parker in 1965 \cite{parker65ssr}, the steady solar wind model only concerns the general dynamical principles but the solar corona is actively heated to maintain the outflow. It is expected that direct observations of the corona and the solar wind will allow more details to be incorporated into the solar wind model. Over the past decades,  significant improvements have been made in solar and solar wind probes, and thanks to these improvements, it has become increasingly clear that the solar wind is indeed not steady and is associated with small-scale impulsive events ubiquitously occurring near the surface of the photosphere. On the theoretical front, with the help of more and more powerful computer simulations, it has become possible to reach a better understanding of the detailed physical processes in the solar corona and how these processes affect the properties of the solar wind \cite{che14prl,che14apjl,kim15apj,vocks16aap,che17pnas,zank17apj,cranmer17ssr,dudik17sp,zank18apj}. 

In this paper, we report some recent progress in the understanding of how the electron beams accelerated by nanoflares shape the solar wind non-thermal electron velocity distribution function (VDF),  generate kinetic waves,  and produce nano-Type III radio bursts. 

\subsection{Evidence for the Connection Between Solar Wind and Nanoflares}
\begin{figure}
\includegraphics[scale=0.8,trim=20 450 10 50,clip]{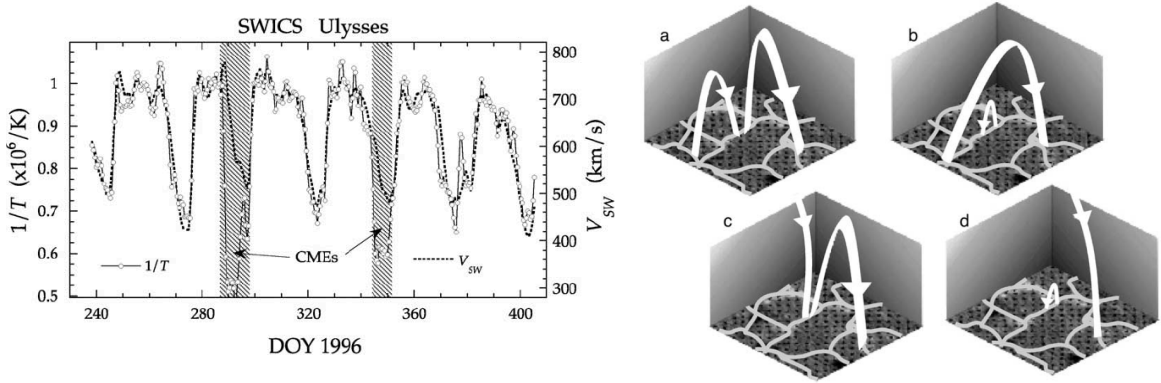}
\caption{{\bf Left:} Time variations of the inverse of the electron temperature $1/T$ (open circles) and of the solar wind proton bulk speed $V_{sw}$ (dotted curves) during a 166-day time period (27 August 1996 to 9 February 1997), observed with the SWICS on Ulysses \cite{gloeckler03jgr}. {\bf Right:} An illustration of the reconnection of loops and open field lines from Fisk (2003) \cite{fisk03jgr}. (a) The foot points of two loops move with convective velocities along the lanes separating the granular and supergranular cells on the solar surface. (b) Two of the foot points of the loops have reconnected to form a new larger loop and a small secondary loop will subduct back into the photosphere. (c) The foot points of a loop and an open field line move along the lanes. (d) A foot point of the loop and the open field line have reconnected, the open field is displaced to lie over the location of another foot point of the loop, and a small secondary loop is again formed that should subduct back into the photosphere.}
\label{fisk}
\end{figure}

\vspace*{0.25cm}
Recent observations of the corona and the solar wind elemental composition and the frozen-in electron temperature strongly suggest that the solar wind is associated with impulsive events close to the surface of the Sun \cite{axford77book,wangym90apj,deforest97sol, zur99ssr,wang03apj,woo04apj,tu05sci,feldman05jgr,brosius14apj,cranmer17ssr}.
It is discovered that 1) the frozen-in electron temperature of the fast wind is $\sim 8\times 10^5$~K, and its nearly photospheric composition suggests that the fast wind may originate from coronal holes; 2) the frozen-in electron temperature of the slow wind is $\sim 1.5 \times 10^6$~K, and its lower coronal composition suggests that slow wind may originate from the quiet Sun. Using data from the Solar Wind Ion Composition Spectrometer (SWICS) onboard {\it Ulysses}  \cite{geiss95sci}, obtained over an entire solar cycle and full latitude range, Gloeckler, Zurbuchen \& Geiss discovered that the solar wind speed is anti-correlated with the electron temperature derived from the density ratio $O^{7+}/O^{6+}$ \cite{gloeckler03jgr}. This discovery implies that the plasma heating process in the lower coronal may affect the acceleration of the solar wind.  Fisk \cite{fisk03jgr,fisk06jgr} proposed that the solar wind was produced by photosphere-rooted small-scale loop-loop magnetic reconnections (MR), and loop-open field-line MR, also known as interchange MR (Fig.~\ref{fisk}). The MR process releases Poynting flux and mass flow that has been heated from the loops into the corona. Assuming a constant mass ejection from the source region, the mass flow escapes to interplanetary along open field lines powered by Poynting flux produces the anti-correlation between the solar wind and the local corona temperature.  

The observations \cite{feldman05jgr,tu05sci} and Fisk's theoretical picture imply that the small-scale impulsive events that occur everywhere in the quiet Sun, including corona holes have similar properties as the nanoflare proposed by Parker \cite{parker88apj}. Recent high-resolution observations of the Sun from sounding rockets, spacecrafts such as SDO, IRIS,  and NuSTAR, are providing a increasingly detailed picture of nanoflares \cite{win13apj,viall13apj,testa14sci,hannah16apj,klim15RSPTA}. The estimated occurrence rate of nanoflares with energy release $\sim 10^{24}$~erg is $\sim 10^6$~$\rm{s}^{-1}$ for the whole Sun.

\subsection{ Coronal Weak Type III Bursts and Nanoflare-Accelerated Electron Beams}
\begin{figure}
\includegraphics[scale=0.8,trim=50 320 60 100,clip]{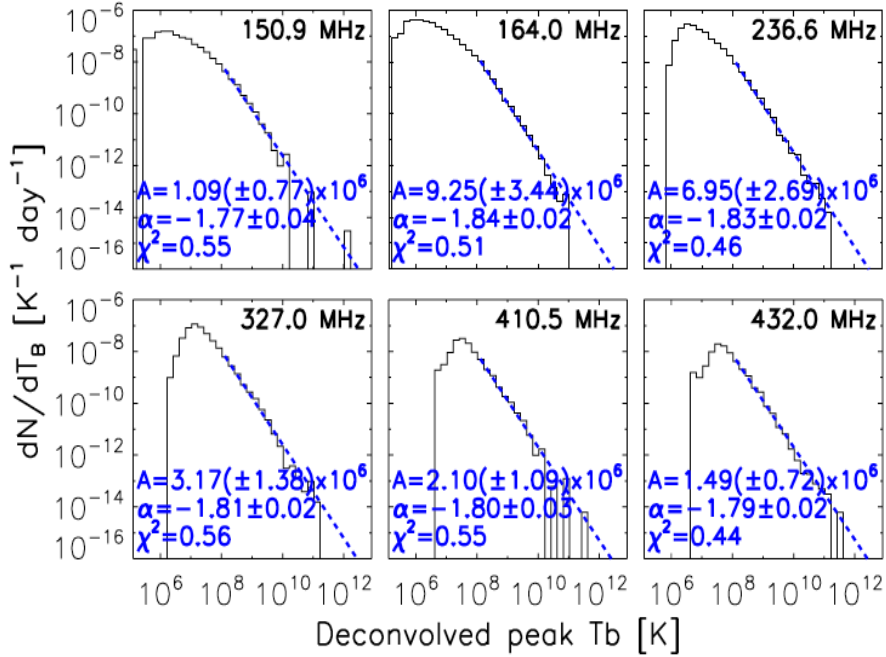}
\caption{ Histogram of deconvolved (i.e., assuming both observed and true sources are Gaussian-shaped) peak brightness temperature ($T_b\sim 10^6$ K) of all type III bursts (The typical $T_b$ of solar flare Type III bursts is about $10^{15}$ K). The dashed blue line is a power-law $dN/dT_b = AT_{\alpha}$
B fit to the data, using the C-statistic (Cash 1979\cite{cash79apj}), technically better suited than Poisson statistics for data sets with
small number of counts per bin (as is our case for the high-value bins, but in this case leading to negligible differences). The associated best-fit parameters are in the lower left corner. Statistical survey of 10,000 type III radio bursts  observed by the Nancy Radioheliograph from 1998 to 2008 found associated with nanoflares (Saint-Hilaire et al, ApJ, 2013 \cite{saint13apj}). }
\label{nano_bt}
\end{figure}

\vspace*{0.25cm}
Similar to flares, nanoflares can accelerate particles and the characteristic energy of nanoflare-accelerated electrons is in keV range \cite{gon13apj}. The accelerated electron beams can trigger electron two-stream instability (ETSI), generate Langmuir waves, and produce type III radio bursts. Indeed, Observations \cite{theja90sol,saint13apj} have found in the solar corona a new kind of type III radio bursts whose brightness temperature $T_b\sim 10^6$ K is about 9 orders of magnitude lower than the flare-associated Type III bursts $T_b\sim 10^{15}$ K and are far more abundant, implying these bursts very possibly originate from nanoflares (see Fig.~\ref{nano_bt}). The high occurrence rate of these ``nano type III bursts" indicates that electron beams and ETSI are common in the solar corona.

\begin{figure}
\includegraphics[scale=0.8,trim=50 350 30 80,clip]{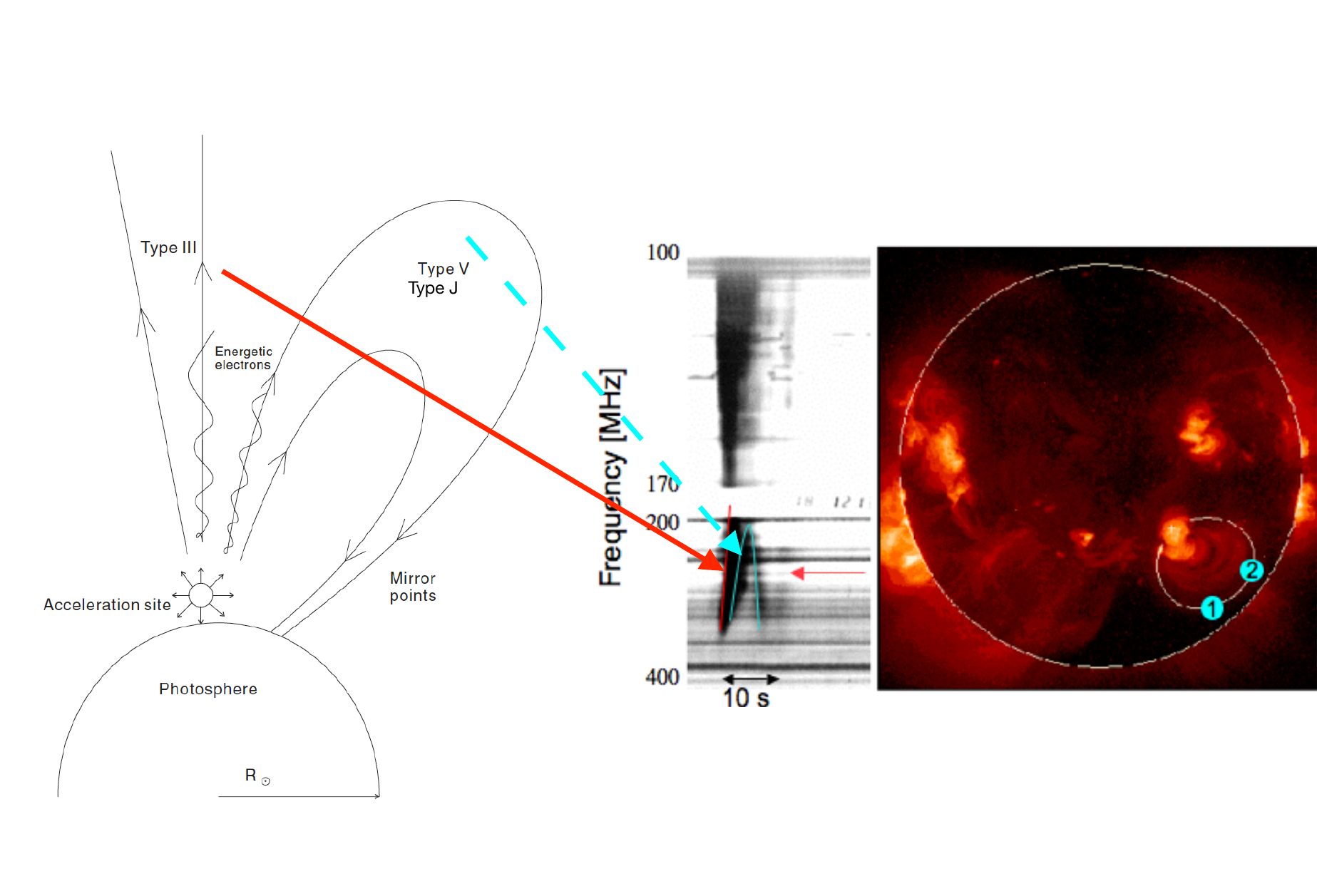}
\caption{{\bf Left}: an illustration of propagation paths for Type III, J and V radio bursts (from Tang et.\thinspace al. 2013 \cite{tang13apj}). {\bf Right}: The spectrum of a meter wave type III burst followed by a Type V burst. The bursts are observed with the \textit{Nançay Radioheliograph}. (Credit: Potsdam Astrophysical Institute.)
}
\label{radio_beam}
\end{figure}

A group of solar radio bursts including Type III, J, and V radio bursts shows the same characteristics: their frequencies are close to the local electron plasma frequency $\omega_{pe}\sim (4\pi n_e e^2/m_e)^{1/2}$, which changes as the electron beams travel along magnetic field lines, as first discovered by wild in 1940s \cite{wild54nat,wild63araa}. Type III radio bursts propagate along open magnetic field lines and can escape from the corona and enter the interplanetary space if the bursts are strong enough, these bursts are called interplanetary bursts; otherwise, bursts are called coronal radio bursts. Type J and V bursts propagate along closed field lines and thus belong to the coronal bursts class (Fig.~\ref{radio_beam}). Nano-Type III radio bursts are weak coronal radio bursts.

If the origin and acceleration of solar wind are associated with the physical processes of nanoflares, the plasma heating produced by electron beams should affect the solar wind properties. Our recent work has addressed this question and is presented in \S~\ref{myres}.
\subsection{Observational Problems of Non-thermal Electron Velocity Distribution Function and Kinetic Turbulence in the Solar Wind}
\begin{figure}
\includegraphics[scale=0.9,trim=90 460 60 10,clip]{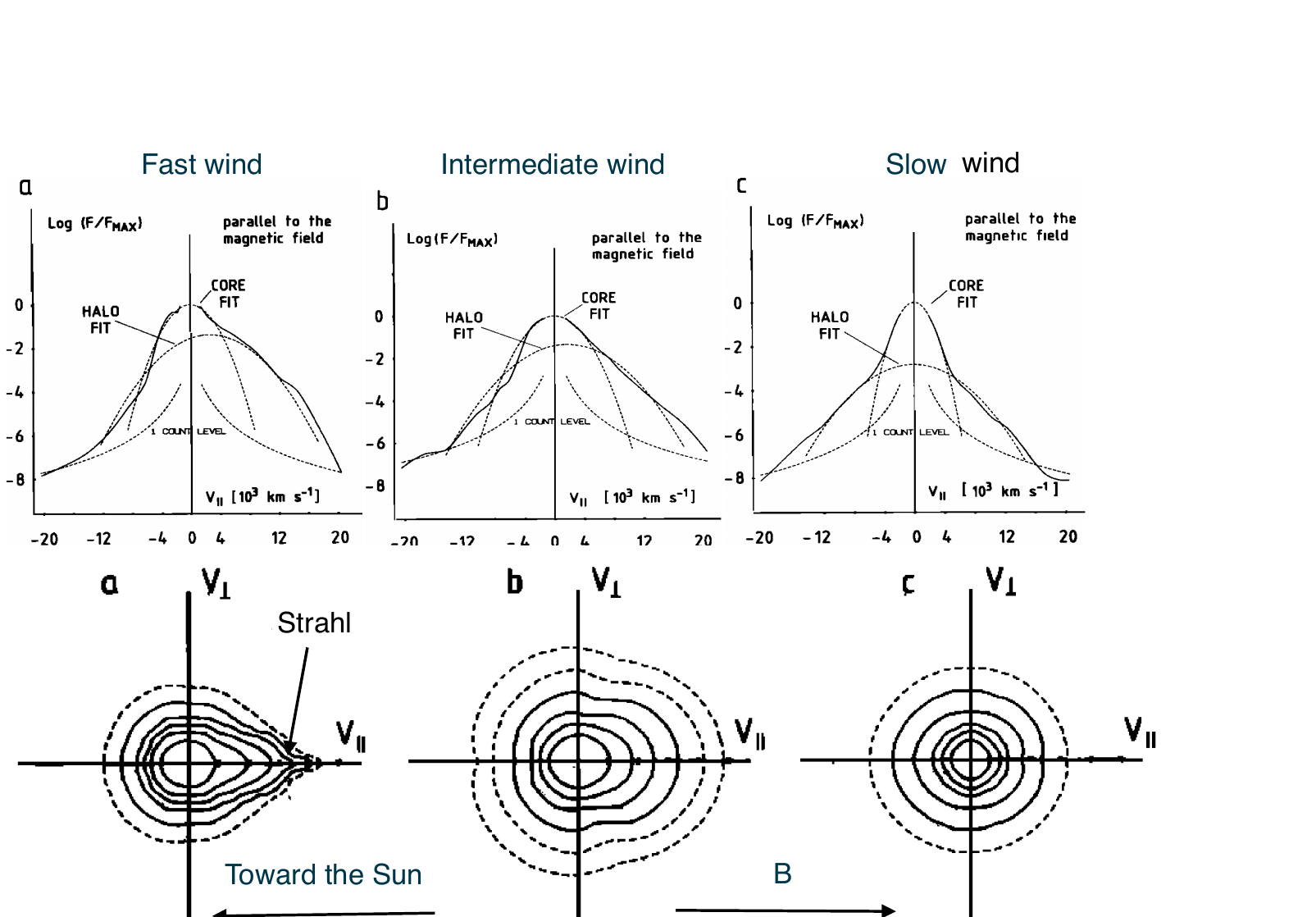}
\caption{The electron VDFs of solar wind at 1AU from Pilipp et.\thinspace al. (1987) \cite{pilipp87jgra}. The top panels show the 1D cuts parallel to the magnetic field crossing the centers of 2D contours displayed below. Slow wind is define with velocity $v_{wind} < 400$~km~s$^{-1}$ and fast wind with $v_{wind} > 700$~km~s$^{-1}$. The speed of the intermediate wind is  between these two. }
\label{pilipp}
\end{figure}

\vspace*{0.25cm}
Observations of electron VDFs at heliocentric distances from 0.3 to 1 AU show a prominent ``break" or a sudden change of slope at a kinetic energy of a few tens of electron volts as shown in Fig.~\ref{pilipp}. The electron VDF below the break is dominated by a Maxwellian known as the {\it core} while the flatter wing above the break is called the {\it halo} \cite{pilipp87jgra,pilipp87jgrb}. How to naturally produce the nearly isotropic halo population that can be described by an approximate Maxwellian function has been a long-standing puzzle in heliophysics \cite{marsch06lrsp}. The isotropic nature of the halo suggests that halo formation needs strong turbulence scattering and is likely related to the kinetic turbulence in the solar wind \cite{marsch06lrsp}. In addition, the {\it strahl} -- an anisotropic tail-like feature skewed with respect to the magnetic field direction is found in the electron VDF of fast solar wind with speed $\sim 400$~km~s$^{-1}$.  In the slow solar wind coming from the sector boundary with speed $< 400$~km~s$^{-1}$, the strahl is nearly invisible and the isotropic core-halo feature dominates. 

Kinetic models show that magnetic focusing effect can produce a strahl-like tail at a minimum heliocentric distance of 10 $R_{\odot}$ \cite{smith_hm12apj,landi12apja}, but the different modes of solar wind turbulence are unable to produce the isotropic halo, suggesting that stronger scattering -- probably caused by kinetic instabilities -- is required \cite{pie99jgr,marsch06lrsp,marsch12ssr,vocks12ssr}. Observations of ion charge states of the solar wind imply a nonthermal tail in the electron VDF of the lower corona, suggesting the coronal origin of the electron halo \cite{ko96grl,esser00apjl,feldman08apj,dzi15apjs}. Whistler waves \cite{vocks03apj,vocks05apj} and other possible kinetic waves \cite{cranmer99apj,laming04apj} in the solar corona are found to be able to effectively scatter the electrons to form the electron halos, but the processes that generate these waves are unknown. 

\begin{figure}
\includegraphics[scale=0.7,trim=50 510 130 10,clip]{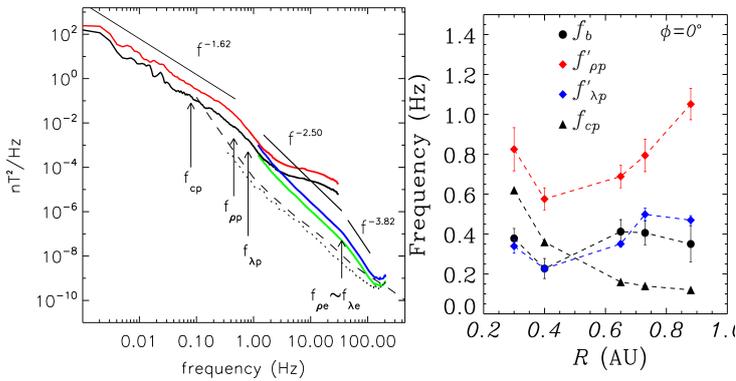}
\caption{{\bf Left:} The observed solar wind magnetic turbulence power spectrum at 1AU \cite{sah09prl}. {\bf Right:} The evolution of magnetic power spectra break from 0.3-1AU \cite{bol12apjl}, where $f_b$ is the observed magnetic power spectral break frequency, $f_{\rho p,e}$ corresponds to the proton (electron) gyro-radius, and $f_{\lambda p,e}$ corresponds to the proton (electron) inertial length. The spectrum is observed in the solar wind co-moving frame.}
\label{break}
\end{figure}

To understand what wave scattering processes produce the observed isotropic electron halo, we need to understand the cause of the kinetic-scale turbulence in the corona and the solar wind. Observations of the solar wind turbulence (Fig.~\ref{break}) have shown that as scales approaching the ion inertial length where wave-particle interactions become important, the power-spectrum of magnetic fluctuations, which in the inertial range follows the Kolmogorov scaling law $B^2_{k} \propto k^{-5/3}$, is replaced by a steeper anisotropic scaling law $B^2_{k_\perp} \propto k_{\perp}^{-\alpha}$, where $\alpha > 5/3$. Spectral index  $\alpha\sim 2.7$ is found in observations but can vary between 2 and 4. Magnetic fluctuations with frequencies much smaller than ion gyro-frequency propagating nearly perpendicularly to the solar wind magnetic field are identified as  kinetic Alfv\'en waves (KAWs) \cite{Leamon_et_al_2000,bale05prl,sah09prl,kiyani09prl,salem12apjl,pod13solphys} and the break frequencies of the magnetic power-spectra from 0.3 to 1 AU suggest the break likely corresponds to the ion inertial length \cite{perri10apjl,bou12apj}. In the past decades, extensive studies of solar wind kinetic-scale turbulence have focused on the idea that the kinetic-scale spectrum is due to the cascade of large-scale turbulence and dissipation on kinetic-scales. However, there are concerns that the energy in the solar wind large-scale turbulence may not be enough to cascade and support the observed kinetic-scale turbulence and heating \cite{Leamon_et_al_1999}. 

In the following section, we show that the electron beams produced by nanoflares can produce the observed core-halo structure and KAWs and whistler waves.

\section{ Electron Two-stream Instability and the Common Origin of Kinetic Turbulence, Nonthermal Electron VDFs and Nano-Type III Radio Bursts}
\label{myres}
\vspace*{0.25cm}
In the classical Kolmogorov turbulence scenario, the energy is injected from the large scale. The balance between energy input and its final absorption is controlled by a nonlinear forward cascade from long wavelengths to dissipation-dominated short wavelengths, resulting in a universal energy cascade power-law with index $-5/3$. In plasmas, the source of instability is often beams of charged particles that inject energy on kinetic scales. For example, electron beams inject energy on electron inertial length $c/\omega_{pe}$ through electron two steam instability (ETSI). Different from Kolmogorov turbulence, at shorter wavelengths the natural candidate to provide the sink of wave energy is Landau damping. However, nonlinear disparate-scale wave interactions which follow from the direct calculation of basic three-wave coupling can only lead to inverse cascades (to longer wavelengths) through modulational
instability \cite{rud78physrep}, and away from the Landau damping region of the spectrum. The eventual nonlinear process capable of overriding this inverse cascade was suggested by Zakharov, namely Langmuir collapse (LC), which is analogous to a self-focusing of the Langmuir wave packets, or cavitons \cite{zak72sjetp}.

How does the ETSI driven by nanoflares shape the kinetic properties of the solar wind?  Our recent particle-in-cell (PIC) simulations clearly demonstrate that the nonlinear effects of ETSI can naturally explain the origin of the observed non-thermal electron VDFs in the solar wind, and the electron beam's contribution to the kinetic scale turbulence is non-negligible \cite{che14prl, che14apjl}. Coherent plasma emission produced by ETSI is found to be able to last for more than five orders of magnitude longer than its linear saturation time, and this long duration possibly resolves the so-called ``Sturrock dilemma" \cite{che17pnas}.  We briefly summarize the major results below.

\subsection{Electron Two-stream Instability}
\vspace*{0.25cm}

ETSI is an electrostatic instability and shows different physical evolutions in cold plasma and warm plasma. The details can be found in a recent review  \cite{che16mpla} and references therein. 

The plasma is cold if the wave phase speed $v_p$ is much larger than the electron thermal speed $v_{te}$, i.e. $v_p \gg v_{te}$.  In the cold plasma limit the phase speed of the fastest growing mode of ETSI is $v_p=(n_b/2n_0)^{1/3}v_d$, where $v_d$ is the drift of electron beams, $n_0$ is the background electron density and $n_b$ is the beam density. In cold plasma ETSI grows with a rate $\gamma\sim \sqrt{3}/2(n_b/2n_0)^{1/3}\omega_{pe}$, and the fastest growing mode is $k_f= \omega_{pe}/v_d$. During the linear growth, most of the kinetic energy of the beams is converted into the growth of electric field $\delta E\propto e^{\gamma t}\sim ek_f m_e (v_{d}-v_{te})^2/2$. The linear growth time-scale $1/\gamma$ is short and comparable to $1/\omega_{pe}$. If $v_d>2v_{te}$ and $\delta E>ek_f m_e v_{te}^2/2$ then the electric field can trap more electrons with velocity $<v_{te}$ and develop electron holes. The trapping and de-trapping of electrons by electron holes can efficiently heat the plasma \cite{che13pop}.

In warm plasma $v_p\sim v_{te}$, the thermal effects must be considered and the kinetic theory is required to describe the ETSI. Different from cold plasma, Langmuir waves are produced and Landau damping becomes the dominant process. Let $\omega=\omega_r +i\gamma$ where $\gamma \ll \omega_r\sim \omega_{pe,0}$, the dispersion relation can be found as for two beams 1 \& 2\cite{bohm49pr, che16mpla}: 
 \begin{eqnarray}
 \label{real} \frac{\omega_r^2}{\omega_{pe,0}^2} &=&1+3k^2\lambda_{D0}^2,\\
\label{im} \frac{\gamma}{\omega_{pe,0}}
 &=&\sqrt{2\pi} [-\frac{\omega_{pe,1}^2 \omega_r}{k^3 v_{t1}^3}e^{-\omega_r^2 / 2k^2 v_{t1}^2}+\frac{\omega_{pe,2}^2 (v_d-\omega_r /k)}{k^2 v_{t2}^3}e^{-(\omega_r/k-v_d)^2/2v_{t2}^2}].
 \label{lw}
  \end{eqnarray}
Eq. (\ref{real}) is the classical dispersion relation of Langmuir wave in a warm
plasma and Eq. (\ref{im}) is the growth rate of Landau damping. 

When the electron beam is strong, ETSI can heat the plasma and cause the plasma to become warm before the kinetic energy in the beam gets exhausted. It is difficult to obtain an analytic solution for what occurs during the transition from cold to warm plasma due to the strong nonlinear effects, particularly, the effects caused by wave coupling and wave-particle interactions. Instead, the complete nonlinear evolution of ETSI is demonstrated using PIC simulations as we describe below.

\subsection{PIC Simulations of Electron Two-stream Instability }
\vspace*{0.25cm}

We have carried out 2.5D massive parallel PIC simulations to study the nonlinear evolution of ETSI in a uniformly magnetized plasma with equal ion and electron temperature \cite{che14prl,che14apjl,che16book}. The initial physical parameters resemble the typical physical condition in the solar corona. The initial density ratio of beam and core is 10\% and the drift of the electron beams is about 10 times larger than the thermal velocity, thus the ETSI starts from a cold plasma and ends as warm plasma. 

\begin{figure}
\includegraphics[scale=0.6,trim=100 60 60 100,clip]{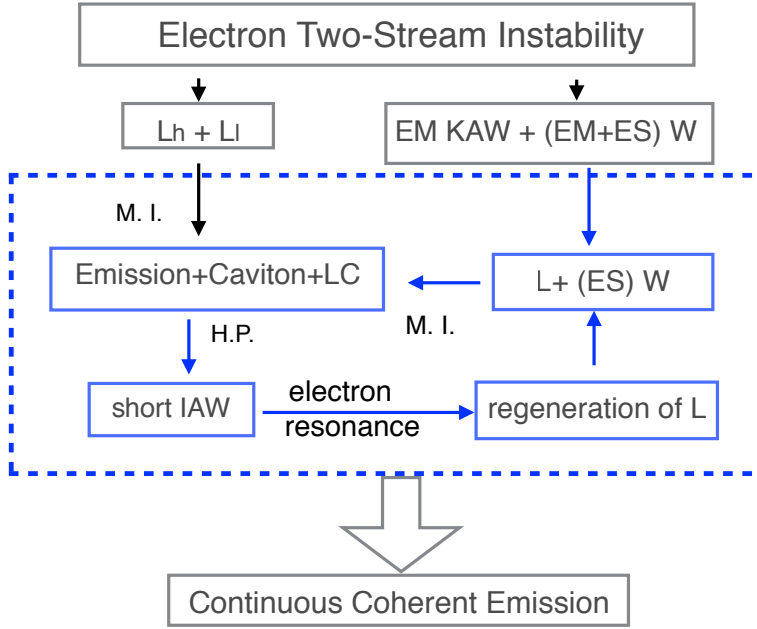} 
\caption{The main processes during the nonlinear evolution of ETSI which includes linear growth, nonlinear growth, saturation and nonlinear decay stages \cite{che17pnas}. The following abbreviations are used: $L_h$: High frequency langmuir wave produced in the background; $L_l$: Low frequency langmuir wave produced inside the electron holes;  EM: Electromagnetic; ES: Electrostatic;  L: Langmuir wave; W: Whistler waves;  EVDF: electron VDF, M. I.: modulational instability; LC: Langmuir Collapse; IAW: ion acoustic wave. Electron holes form in the nonlinear growth stage. Modulational instability, which occurs at the saturation stage through the  nonlinear decay stage, leads to Langmuir collapse and electron heating that fills in cavitons. The high pressure is released via the excitation of a short wavelength IAW that is damped by electrons and re-excites small-scale Langmuir waves---this process closes a feedback loop (outlined by blue dashed lines) that maintains the continuous coherent emission.  }
\label{mainchart}
\end{figure} 

The ETSI experience four phases: linear growth, nonlinear growth, saturation and nonlinear decay till turbulent equilibrium. The time-scale of the linear growth phase is about tens of $\omega_{pe}^{-1}$ while the time-scale of the nonlinear evolution of ETSI is $\sim 10^4 \omega_{pe}^{-1}$. The complete evolution of ETSI is shown in Fig.~\ref{mainchart}. The linear growth stage of ETSI can be well described by the cold plasma limit. Then electron holes form in the nonlinear growth stage and produce fast electron heating, the plasma becomes warm and enters the nonlinear evolution stage. Two parallel processes are found:  the generation of low frequency KAWs and whistler waves through bi-directional energy cascades,  and the generation of high-frequency Langmuir waves. The plasma emission is continuously maintained through the repeating modulational instability driven by the disparate-scale wave coupling between Langmuir waves and low-frequency waves (shown in the blue box in Fig.~\ref{mainchart}). The wave-wave and wave-particle interactions eventually lead to the balance of energy exchange, and the turbulence stays at a nonthermal equilibrium in which the non-Maxwellian electron VDF and kinetic waves co-exist, a self-consistent solution of Vlasov equation \cite{kim15apj}. These relics of the violent dissipation through ETSI can be carried out into interplanetary with the solar wind along open field lines \cite{che14apjl}. 

\subsection{ETSI and the Common Origin of Kinetic Turbulence and Non-thermal Electron VDFs}
\label{ETSI1}
\vspace*{0.25cm}
ETSI injects beam kinetic energy on scales close to the Debye length, and the energy cascade is quickly stopped by strong electron heating. On the other hand, we found that the coupling between KAW and whistler waves can inversely transfer the energy to large scales and develop kinetic-scale turbulence. Simultaneously, the waves scatter the hot electron tail that lies along the magnetic field into an isotropic population superposed over the Maxwellian core electron VDF, forming the electron halo in the solar wind \cite{che14prl,che14apjl}.
\begin{figure}
\includegraphics[scale=0.8,angle=0,trim=70 500 80 50,clip]{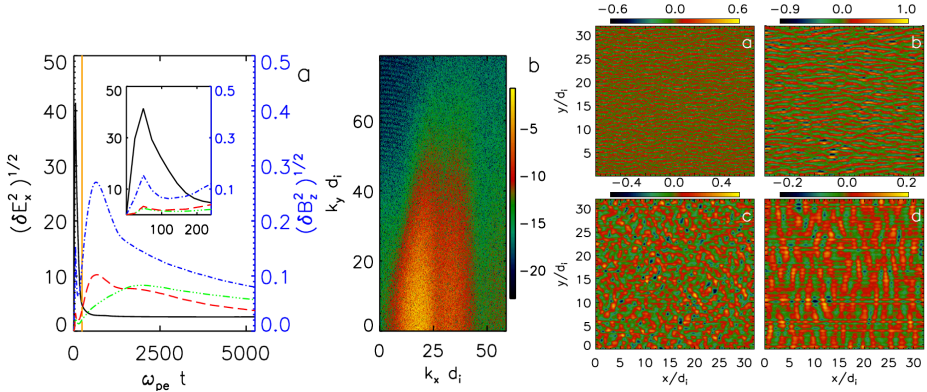} 
\caption{\textbf{Left:} Time evolution of turbulent energies in the PIC simulation: $\langle \delta E_x^2\rangle$ (black line), $\langle \delta B_x^2\rangle$ (red line), $\langle \delta B_y^2\rangle$ (green line), $\langle \delta B_z^2\rangle$ (blue line). The embedded is an expanded view of the time evolution from $\omega_{pe} t=0-230$. The orange line indicates $\omega_{pe} t=230$. \textbf{Middle:} Power spectrum of $\vert \delta E_x(k_x, k_y)\vert^2$ at the linear phase of ETSI, shown in logarithmic scale. \textbf{Right:} $B_z/B_0$ at ({\bf a}) the linear phase, ({\bf b}) the nonlinear growth phase,  ({\bf c}) the nonlinear decay phase, and ({\bf d}) the turbulence equilibrium. (reference by Che et al \cite{che17prl})}
\label{fluex}
\end{figure} 

The linear growth of ETSI lasts about $\omega_{pe}t=20$ (Fig.~\ref{fluex}), and  $\sim 10\%$ of the kinetic energy in the beam is converted into magnetic energy at the nonlinear growth phase, while nearly 90\% is converted into the thermal motion of trapped electrons. The fast growth of the electric field induces a magnetic field, and the electric current density  $j_{ex}$ produced by the inductive magnetic field becomes as important as the displacement current when the ETSI starts to decay. The current $j_{ex}$ then drives a Weibel-like instability that generates nearly non-propagating transverse electromagnetic waves. The fast decay of the localized $j_{ex}$ breaks up the transverse waves and produces randomly propagating KAWs and whistler waves. The wave-wave interactions drive a bi-directional energy cascade. The perpendicular KAW energy is transferred from the electron inertial scale up to the ion inertial scale. The parallel whistler wave energy is transferred from the ion inertial scale down to the electron inertial scale. Eventually, magnetic power is concentrated in two branches in the energy spectrum: the nearly perpendicular branch with $k_x d_i<1$, and the parallel branch with $k_y d_i <2$ (Fig.~\ref{fluex}). Around $\omega_{pe} t=10000$, the energy exchange between particles and waves reaches a balance. The turbulence reaches its new steady state with $P^2 + B^2/8\pi=constant$, where $P$ is the total pressure of ions and electrons.
\begin{figure}
\includegraphics[scale=0.8,angle=0,trim=70 550 80 50,clip]{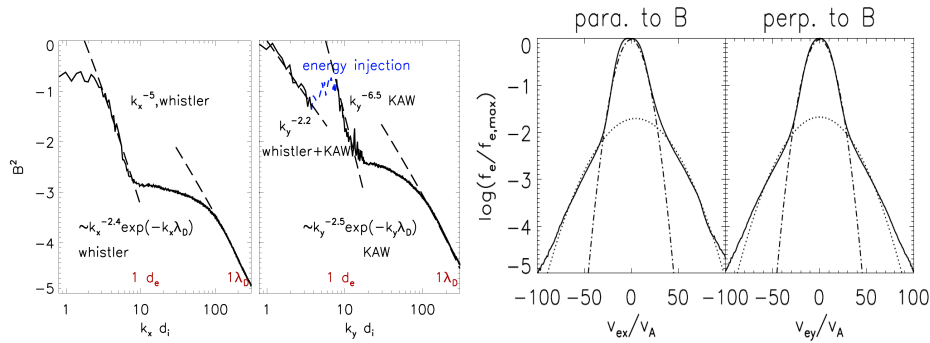} 
\caption{\textbf{Left:} The 1D spectra of $ \delta B^2(k)$ in x  and y directions \cite{che14prl}. The blue short-dashed line shows the wavenumber where magnetic energy injection occurs. \textbf{Right:} The 1D electron VDF cuts parallel and perpendicular to magnetic field when ETSI simulation reaches turbulence equilibrium \cite{che14apjl}. The dot-dashed lines delineate the core Maxwellian VDF and the dashed lines represent the halo VDF. They are plotted in the same manner as in Pilipp et. al. (1987) \cite{pilipp87jgra}.  }
\label{spec1d}
\end{figure} 

The amplitude ratio of the magnetic fluctuations to the background magnetic field is $\sim 0.2$, which agrees with observations of solar wind kinetic turbulence. The 1D magnetic fluctuation power spectrum is shown in Fig.~\ref{spec1d}. With $1<k_y d_i<2$, which corresponds to the range of wavelengths current instruments can probe, both KAWs and whistler waves are important. The perpendicular power spectrum is fitted with a power-law with an index of -2.2 which falls within the observed range. The perpendicular power spectrum terminates at the ion inertial length, is also consistent with observations \cite{Leamon_et_al_1999,perri10apjl,bou12apj}. A ubiquitous observable feature is a spectral break at the electron scale caused by energy injection. This model also predicts the existence of whistler waves, and the cutoff of the parallel power spectrum at the ion gyro-radius \cite{che14prl}. 

The steady-state electron VDF in our simulation agrees with the observed core-halo structure in the solar wind (Fig.~\ref{spec1d}). This is expected if the beam heated plasma escapes from the inner corona and advects into interplanetary medium along open field lines, forming the solar wind and preserving its kinetic properties. This nonlinear heating process predicts that the core-halo temperature ratio $T_h/T_c$ of the solar wind is insensitive to the initial conditions in the corona but is linearly correlated to the core-halo density ratio of the solar wind $n_c/n_h$: 
\begin{equation}
\frac{T_h}{T_c}  \approx \frac{n_c}{n_h}\frac{1-C_T}{C_T}+4,
\label{dthc2}
\end{equation} 
where $C_T$ is the rate at which kinetic energy of electron beams converts to heat, and $C_T \sim 0.9$ is found in our simulations. If the core and halo experience similar temperature evolutions when traveling from the Sun to 1AU, the temperature ratio can be approximately preserved.  In fast wind where the strahl is strong,  the halo temperature can be replaced by the mean temperature of halo and strahl $T_{hot}=n_{strahl} T_{strahl}/n_c +n_h T_h/n_c$ and halo density be replaced by the total density of both strahl and halo $n_{hot}=n_{strahl}+n_h$ because the energy and density are approximately conserved during scattering \cite{mak05jgr}. Thus we have:
\begin{equation}
\frac{T_{hot}}{T_c}  \approx \frac{n_c}{n_{hot}}\frac{1-C_T}{C_T}+4,
\label{dthc3}
\end{equation} 
The break point dividing the core and halo in electron VDF, which is a useful quantity in observations, satisfies:
\begin{equation}
v_{brk}\approx[\ln(T_{h}/T_c)-\ln (n_h/n_c)^2]^{1/2} v_{te,c}.
\label{brk}
\end{equation} 
In addition, the relative drift between the core and halo is close to the core thermal velocity -- a relic of the ETSI saturation.

Our simulations \cite{che14prl} show that when the kinetic turbulence fully saturates, the ratio of parallel to perpendicular electric field fluctuations $\langle \vert \delta E_{\parallel}\vert/\vert \delta E_{\perp}\vert\rangle$ is enhanced by the relic parallel electric field by a factor of $\sim 2-3$, consistent with observations that the parallel turbulent electric field is larger than the perpendicular turbulent electric field but contrary to what is expected if the turbulent fluctuations are dominated by KAWs \cite{mozer13apjl}. The enhanced electric field might be caused by electrostatic whistler waves and Langmuir waves. 

\subsection{ETSI and  Continuous Coherent Plasma Emission }
\label{ETSI2}
\vspace*{0.25cm}

The continuous plasma coherent emission is produced while the electron halo and the KAWs and whistler waves develop as have been shown in detail in Che et.\thinspace al. (2017) \cite{che17pnas}. Here we present a basic picture. 

Ginzburg and Zhelezniakov \cite{ginzburg59book} in 1958 first proposed a basic framework for Type III radio bursts. The essence of the scenario is that the ETSI, driven by electron beams, generates Langmuir waves that are converted into plasma coherent emission via nonlinear three-wave coupling (e.g., One Langmuir wave, one ion acoustic wave (IAW) produce one photon).  However, the deacceleration time of ETSI in solar corona is $\sim 10^{-7}$s estimated by the growth rate in warm plasma shown in Eq.~\ref{lw}\cite{sturrock64}, which is more than five orders of magnitude shorter than the duration of radio bursts. This long-standing problem was first pointed out by Sturrock and is known as the ``Sturrock's dilemma"  \cite{sturrock64}.  

Several theoretical models have been proposed to refine the Ginzburg \& Zhelezniako model to address this problem \cite{papa74apj,goldstein79apj,Freund80pof,goldman83sol,
robinson97rmp}. However, there are two major problems in these models. 1) All the models focus only on the regeneration of Langmuir waves and ignored the generation of the low-frequency waves. These models assume that IAWs are present in the background which is not always true in the realistic environment. On the other hand, the coupling between the two waves requires the waves to be in phase, a condition that cannot be guaranteed if the IAWs are not self-consistently produced in the process. 2) All the models are based on quasi-linear theory. But it is found that Langmuir collapse is often associated with Type III radio bursts \cite{robinson97rmp,reid14raa}. Langmuir collapse is a strong turbulence process, which contradicts to the conditions required for the quasi-linear theory, implying the regenerated Langmuir waves by these model must be much weaker than what occurs in the observations.  With PIC simulations we show our strong ETSI model can overcome these two problems and provide a self-consistent mechanism to continuously generate coherent radio emission.  

In the aforementioned PIC simulation of the ETSI, the coherent plasma emission is not present until the saturation stage when the plasma becomes warm due to the heating by electron holes. At this stage, $\omega_{pe,0}t=320-420$, the high-frequency Langmuir wave $L_h$ is generated in the background plasma and the low-frequency Langmuir wave $L_l$ is generated in the trapped electrons inside the electron holes. These two Langmuir waves satisfy the following dispersion relation (normalized by the initial $\omega_{pe,0}$ and $\lambda_{De,0}$):
\begin{equation}
\frac{\omega}{\omega_{pe,0}}=\left ( \frac{n_e^2}{n_0^2}+\frac{T_{ce} n_e}{T_{ce,0} n_0} \gamma k_x^2\lambda_{De,0}^2\right )^{1/2},
\label{lw1}
\end{equation}
where $\gamma = 3$ as the electron heating caused by the solitary wave is nearly adiabatic \cite{che13pop}.

\begin{figure}
\includegraphics[scale=1,angle=0,trim=40 560 0 70,clip]{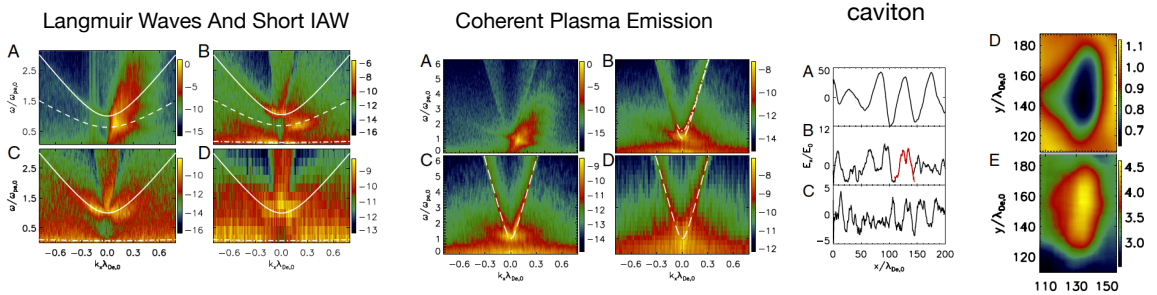} 
\caption{{\bf Left}: The $\omega/\omega_{pe,0}-k_x \lambda_{De,0}$ diagrams of the parallel propagating high frequency electric field component $E_x$ at four time intervals:  (A): $\omega_{pe,0}t=0-100$; (B): $\omega_{pe,0}t=320-420$; (C): $\omega_{pe,0}t=2880-2980$; and  (D): $\omega_{pe,0}t=10560-10580$. Also shown are the Langmuir wave dispersion relations of the background electrons (solid lines), trapped electrons (dashed lines), and the short wavelength IAW (dash-dotted lines). {\bf Middle}: The  $\omega/\omega_{pe,0}-k_x \lambda_{De,0}$ diagrams of parallel propagating high frequency $E_y$ are shown for the same four time intervals for Langmuir waves. Dashed lines: dispersion relation of plasma emission with base frequency $\sim \omega_{pe,0}$.  {\bf Right}: Parallel electric field $E_x$ for $x \in [0, 200\lambda_{De,0}]$ and $y=100 \lambda_{De,0}$ at (A) $\omega_{pe,0} t=72$ when the ETSI nearly saturates and hot electrons excite Langmuir waves, (B) $\omega_{pe,0} t=320$ when the modulational instability grows, LCs start and cavitons form, and (C)  $\omega_{pe,0} t=648$ LCs continue.  The ion density map (D) and the electron temperature map (E) of the caviton that corresponds to the Langmuir envelope (in red) in panel (B) are also shown. Figures from the paper by Che et al.\cite{che17pnas}.}
\label{emission}
\end{figure}

The coupling between the two Langmuir waves $L_h+L_l\rightarrow T$ (where $T$ is transverse emission) drives modulational instability and produces the first coherent emission with frequency about $1.6\omega_{pe,0}$ (Fig.~\ref{emission}). The emission propagates forward much stronger than backward and satisfies the dispersion relation:
 \begin{equation}
\frac{\omega}{\omega_{pe,0}}=\left (\frac{n_e^2}{n_0^2} +\frac{c^2}{v_{te}^2}k_x^2\lambda_{De,0}^2\right )^{1/2}.
\label{light}
\end{equation}

The maximum growth rate for modulational instability is $\gamma_{m}=\omega_{pi}(\langle E_L^2\rangle/8\pi n_0 T_e)^{1/2}$ and the critical condition for Langmuir collapse (LC) is \cite{zak72sjetp}:
\begin{equation}
\frac{E_L^2}{8\pi n_e T_e}>\frac{1}{4}k_x^2\lambda_{De}^2.
\label{critic}
\end{equation} 
 
As the modulation instability grows and the critical condition in Eq.~(\ref{critic}) is satisfied, Langmuir collapse (LC) occurs \cite{zak72sjetp,rud78physrep}. LC leads to the contraction of the modulated Langmuir envelope and the formation of ion density cavitons. We plot an example of the parallel electric field $E_x$  in Fig.~\ref{emission} at three moments: $\omega_{pe,0} t=$ 72, 320, 680. At $\omega_{pe,0} t=$ 72, the solitary waves with wavelengths near the fastest growing mode reach the peak. The critical condition for LC $E^2/8\pi n_0 T_e>\frac{1}{4}k_x^2\lambda_{De}^2$ is satisfied since $(E^2/8\pi n_0 T_e)^{1/2}\sim 0.4$ is larger than the fastest growing mode of the ETSI  $k\lambda_{De}/2\sim v_{te,0}/2v_{db,0}\sim 0.05$.  At $\omega_{pe,0} t=$ 320, the modulated wave envelopes decrease from 50 to 30~$\lambda_{De,0}$ and ion density cavitons form. Contraction of the Langmuir wave envelopes efficiently dissipates the Langmuir wave energy into electron thermal energy. 

Consequently, LC results in the destruction of the cavitons and the release of hot plasma that was inside the caviton. The process generates intermediate short ion acoustic waves (IAWs).  Short IAWs resonate with the electrons and regenerate  Langmuir waves. The emission is maintained by the cyclic couplings between the regenerated Langmuir waves and the electrostatic component of the whistler waves (Fig.~\ref{mainchart}). The frequency of the intermediate short IAW is of the same order as the ion plasma frequency $\omega_{pi}$, and satisfies a dispersion relation similar to that of Langmuir waves:
 \begin{equation}
\omega^2\approx \omega^2_{pi}\frac{1+\sqrt{3T_i/T_e}}{4} + \sqrt{\frac{3T_e}{T_i}}k^2 v_{ti}^2.
\label{ia}
\end{equation} 

The Langmuir waves, short IAWs, and emissions at four different evolution stages are shown in Fig.~\ref{emission}. For each cycle, the regeneration of the Langmuir wave leads to a small frequency shift and a reduced amplitude of the Langmuir waves.  The emission continues beyond the saturation of turbulence.

In our simulations, the ETSI nonlinear saturation time is $\sim 1.5\times 10^4\omega_{pe}^{-1}$. Since the modulational instability nearly dominates the entire process, the nonlinear saturation time is approximately proportional to $(m_i/m_e)^{1/2}$, and for real mass ratio, the ETSI nonlinear saturation time is translated to $\sim 10^5 \omega_{pe}^{-1}$, which is significantly longer than the ETSI linear saturation time $(n_0/n_b)^{1/3} \omega_{pe}^{-1}\sim 2 \omega_{pe}^{-1}$. Note that our simulation assumes instantaneous injection of the electron beam, while in the corona the electron-acceleration time is finite and the beam will propagate out of the region of initial generation. The acceleration time also affects the actual duration of the bursts \cite{goldstein79apj,rat14aap}. The overall scenario is that coronal bursts produce nonthermal electrons that escape into space and produce interplanetary bursts \cite{aschw02ssr} with accompanying waves. Our simulation assumes the beam energy is about 100 times the coronal thermal energy. For nanoflares, the beam energy is about 1 keV and the corona temperature is $\sim10$ eV.  We estimate the emission power in our simulation is $\sim 10^{-4}-10^{-6}$ of the Langmuir wave power. Such small energy loss is negligible dynamically. The emission mechanism we discussed provides a self-consistent solution to the long-standing ``Sturrock's dilemma" \cite{sturrock64}. In Table~1 we provide an incomplete sample of literature in which the observations of various types of radio bursts are consistent with our model predictions.
\begin{table}
\centering
\caption{Model Predictions and Observational Evidence}
 \resizebox{0.9\textwidth}{!}{%
 \begin{tabular}{c | c | c}
 \hline\hline 
 Model Predictions & Observations & References \\
 \hline
 In the solar corona emission duration  & Coronal Type J \& U radio bursts & \cite{aschw95apj,aschw02ssr,saint13apj}  \\
 $\sim 10^5 \omega_{pe}^{-1} \sim 1-10$~ms. &
  Weak Coronal Type III radio bursts &   \\
 \hline
 Langmuir waves \& whistler waves & Interplanetary Type III radio bursts & \cite{lin86apj,kellogg92grlb,mac96aa,ergun08prl} \\
 \hline
 Langmuire collapse \& short wavelength IAW & Interplanetary Type III radio bursts & \cite{lin81apj,lin86apj,kellogg92grla}\\
 
  \hline\hline 
\end{tabular}}
\end{table}

\section{Summary and Open Questions}
\vspace*{0.25cm}
To summarize, we have shown that nanoflare-accelerated electron beams can trigger ETSI, which generates kinetic turbulence as well as a non-Maxwellian electron VDF, consistent with observations of the solar wind \cite{che14apjl}. The major attraction of this finding is that it can account for the origin of both the electron VDF and kinetic turbulence in a unified picture, while past studies treat these two phenomena as unrelated. The link between the solar wind and nanoflares directly relates solar wind properties to photospheric dynamics and puts useful constraints on kinetic processes in both the solar corona and the solar wind. The plasma coherent emission produced in our model agrees well with the radio observations of nano-Type III, J \& V solar radio bursts. The model also predicts features that can be tested with current and future solar and solar wind probes. One of the most important predictions of our model is the correlation between the temperature of core and halo of solar wind electron VDF, and this correlation is confirmed in a recent analysis of 12 years of \textit{WIND} data \cite{macneil17ag}.  Recent SDO observations of the corona also suggest that the plasma heating is associated with open and close field \cite{orange16apj}. Using CHIANTI data base, it is found electron VDF is non-Maxwellian \cite{dzi15apjs} and electron beams form in the lower corona \cite{dz16aap}. 

The upcoming Parker Solar Probe (PSP) and Solar Orbiter (SO) spacecrafts will provide unprecedented \textit{in situ} observations of the solar wind and multi-bands remote observations of corona activities. The \textit{in situ} observations of the solar wind around 10 solar radii can test whether the electron halo and strahl in the electron VDF are of coronal origin \cite{cranmer17ssr,dudik17sp,graham18apj}. The observations from 10 solar radii to 1AU will enable us to investigate the evolution of electron and ion VDFs in the inner heliosphere. The simultaneous X-ray and radio observations will provide us more information on the particle acceleration in solar flares. 

How the particle heating and acceleration in the corona affect the properties of the solar wind is the core science for both PSP and SO. Nanoflares \cite{benz04iau,testa14sci,klim15RSPTA} and plasma waves are two dominant sources of coronal heating \cite{pon14sci,vanb17apjb,zank17apj,zank18apj}. Therefore, {\it in situ} observation of the solar wind opens a window to nanoflare heating.

There are several unsolved problem related to nanoflares induced particle energization:  1) MR is believed to be the engine of particle energization in nanoflares. The interchange MR is essential for slow solar wind to escape from the lower corona to interplanetary space. How MR accelerates and transports particles in the solar corona is still not very well understood and is a subject of active study \cite{drake06nat,zank14apj,zank15apj,guo16apjl,boro17apj,dahlin17pop}.  Current observations \cite{aschw05book,benz17lrsp} and theoretical/numerical models \cite{antiochos07apj,antiochos11apj,higg17apjl} of interchange MR still cannot provide sufficient details on how beams and energetic particles are produced; 2) The ion VDF of the solar wind is also non-Maxwellian \cite{marsch82jgr}. How the ion beams in the corona impact the solar wind ion VDF has not been closely investigated;  3) Energetic particle propagation is very important in space weather applications. In the past, the study of energetic particle transport in the heliosphere focuses on particle scattering and acceleration by Alfv\'enic wave turbulence, and the role of kinetic scale turbulence on solar wind has not drawn sufficient attention, particularly the impact on solar wind electrons such as the evolution of the anisotropic strahl \cite{graham18apj}. 

How to incorporate kinetic processes into the global solar wind model is a profound theoretical and observational challenge. The study of the impact of nanoflares on the solar wind's properties on both large and kinetic scales will certainly be enlightening to the pursuit of a complete understanding of the origin and interactions of the solar wind. Such understanding will have broader implications for astrophysical winds and outflows beyond the solar system. 

\section{Acknowledgement}
HC would like to thank all collaborators on this project:  Drs. M. Goldstein, P. Diamond, R. Sagdeev and A. Vin\~as.  HC would like to thank Dr. G. Zank for the kind invitation to the {\it 17th Annual International Astrophysics Conference} and inspiring discussions during the meeting. HC also like to thank the participants at this meeting for the interesting discussions. HC is partly supported by NASA Grant No.NNX17AI19G. The simulations and analysis were supported by the NASA High-End Computing (HEC) Program through the NASA Advanced Supercomputing (NAS) Division at Ames Research Center.
\section*{References}
\vspace*{0.25cm}
\bibliographystyle{iopart-num}

\providecommand{\newblock}{}

\end{document}